\documentclass[
 reprint,
 amsmath,amssymb,
 aps,
 prl
]{revtex4-1}

\usepackage{times}
\usepackage{graphicx}
\usepackage{dcolumn}
\usepackage{bm}
\usepackage{mathtools}
\usepackage[english]{babel}
\usepackage{physics}
\usepackage{enumitem}
\usepackage{amsmath}
\usepackage{mathrsfs}
\usepackage{xfrac}
\usepackage{xr-hyper}
\usepackage{hyperref}
\usepackage{footmisc}
\usepackage{hypcap}
\usepackage{makecell}
\usepackage{url}
\usepackage{ifsym}

\usepackage{bbold}
\usepackage{gensymb}
\usepackage{multirow}
\usepackage{array}
\usepackage[draft]{changes}
\definechangesauthor[name=Guilherme,color=green]{GZ}
\definechangesauthor[name=Maxime,color=blue]{MJ}
\definechangesauthor[name=Lee,color=red]{LR}

\renewcommand{\ket}[1]{|#1\rangle}
\renewcommand{\bra}[1]{\langle#1|}

\begin{document}
\title{Fiber-compatible photonic feed-forward with 99\% fidelity}\

\author{Guilherme Luiz Zanin$^{1,2,\ast\dagger}$}
\author{Maxime J Jacquet$^{1\dagger}$}%
\author{Michele Spagnolo$^1$}
\author{Peter Schiansky$^1$}
\author{Irati Alonso Calafell$^1$}
\author{Lee A Rozema$^1$}
\author{Philip Walther$^{1,3}$}

\affiliation{
$^1$Vienna Center for Quantum Science (VCQ), Faculty of Physics, University of Vienna, Boltzmanngasse 5, Vienna 1090, Austria\\
$^2$Departamento de F\'{i}sica, Universidade Federal de Santa Catarina, Florian\'{o}polis, Santa Catarina 88040-900, Brazil\\
$^3$Christian Doppler Laboratory for Photonic Quantum Computer, Faculty of Physics,  University of Vienna, 1090 Vienna, Austria\\
$^\ast$Correspondence to:  guilherme.zanin@univie.ac.at.\\
$^\dagger$These authors contributed equally to this work.
}

\date{\today}

\begin{abstract}
Both photonic quantum computation and the establishment of a quantum internet require fiber-based measurement and feed-forward in order to be compatible with existing infrastructure.
Here we present a fiber-compatible scheme for measurement and feed-forward, whose performance is benchmarked by carrying out remote preparation of single-photon polarization states at telecom-wavelengths.
The result of a projective measurement on one photon deterministically controls the path a second photon takes with ultrafast optical switches.
By placing well-calibrated {bulk} passive polarization optics in the paths, we achieve a measurement and feed-forward fidelity of (99.0 $\pm$ 1)\%,  after correcting for other experimental errors.
Our methods are useful for photonic quantum experiments including  computing, communication, and teleportation.
\end{abstract}

\maketitle

\section{Introduction}
Linear optical quantum computing (LOQC) \cite{kok_linear_2007} has gained tremendous momentum since its inception in the early 2000s, when Knill, Laflamme and Milburn (KLM) showed that it is possible to implement the two-photon interactions necessary for quantum computation by means of post-selection and ancilla photons, thus producing near-deterministic gates with a scalable polynomial overhead \cite{knill_scheme_2001}.
Although the gate-based approach to photonic quantum computing suffers from an unsustainable overhead that is polynomial in the asymptotic limit only \cite{hayes_utilizing_2004},
there exists an alternative universal quantum computing paradigm known as `one-way' \cite{yoran_deterministic_2003,raussendorf_one-way_2001,walther_experimental_2005} or `measurement-based' quantum computation (MBQC) \cite{raussendorf_measurement-based_2003}.
MBQC uses an entangled multiparticle state, the cluster or graph state, as an input \cite{nielsen_optical_2004}.
Optical cluster states can be efficiently created by the Browne-Rudolph fusion mechanism \cite{browne_resource-efficient_2005}, and it is even possible to renormalize an imperfect cluster state by applying ideas of percolation theory \cite{kieling_percolation_2007}.
This is now widely considered to be the most promising route to photonic quantum computing \cite{rudolph_why_2017}.
{In any case, LOQC using the circuit model, MBQC, constructing a quantum network \cite{kimble2008quantum}, and many other important quantum protocols, including quantum teleportation \cite{giacomini_active_2002,ursin_quantum_2004,valivarthi2020teleportation}, remote state preparation \cite{bennettRSP}, quantum error correction \cite{vijayan2020robust}, improving photon sources \cite{kanedaHighefficiencySinglephotonGeneration2019a,meyer-scottExponentialEnhancementMultiphoton2019}, and even quantum metrology \cite{higgins2007entanglement} and quantum data compression \cite{rozema2014quantum}, require one to conditionally apply a quantum operation based on the outcome of an earlier measurement \cite{pittman_demonstration_2002,prevedel_high-speed_2007,saggio_experimental_2019}.
Given the extremely wide range of applications, effective feed-forward techniques will be required in bulk optics, integrated optics and fiber optics.
Here we present a new technique for feed-forward, and apply it in a experiment.


Measurement and feed-forward has historically been done in bulk optics by means of Pockels cells {\cite{giacomini_active_2002,walther_experimental_2005,sciarrinoRealizationMinimalDisturbance2006,bohi_implementation_2007,ma_experimental_2012}}, whose tunable retardance directly implements an operation on the polarization state of single photons.
However, each Pockels cell can only rotate about one axis on the Bloch sphere; therefore, realizing a general unitary transformation requires the use of 3 active elements, each of which needs to be precisely calibrated.
{When it comes to quantum information processing, this} results in large technical overheads.
As such, even experimental demonstrations, whose goal is to develop real-world deployable technologies, often use post-selection rather than real feed-forward \cite{valivarthi2020teleportation}.

Here, we {explore an alternative approach to measurement and feed-forward with bulk components: we route} photons through passive elements using simpler active elements, \textit{i.e.} ultrafast optical {(electro-optic)} switches (UFOS).
The critical component is the switch that receives a signal from the heralding detector, as this redirects the heralded photon towards the polarization correcting components.
Recent schemes have used integrated switches, such as fast opto-ceramic switches \cite{xiongBidirectionalMultiplexingHeralded2013,collinsIntegratedSpatialMultiplexing2013,hoggarthResourceefficientFibreintegratedTemporal2017}, electro-optic switches \cite{bridaExperimentalRealizationLownoise2011,bridaExtremelyLownoiseHeralded2012,meanyHybridPhotonicCircuit2014,mendozaActiveTemporalSpatial2016,francis-jonesFibreintegratedNoiseGating2017,massaro_improving_2019} or bulk electro-optic polarization rotating switches \cite{jeffreyPeriodicDeterministicSource2004,ma_experimental_2011,pittman_demonstration_2002,kiyoharaRealizationMultiplexingHeralded2016,kaneda_time-multiplexed_2015} or phase modulators \cite{mikovaIncreasingEfficiencyLinearoptical2012,mikovaOptimalEntanglementassistedDiscrimination2014,mikovaFaithfulConditionalQuantum2016} in interferometers .
In all cases, the heralded photons must be delayed to allow time to process the heralding signals and activate the switch.
The ensuing latency is largely dominated by electronic processing time, with typical latency values ranging from 20 to 1000 ns \cite{giacomini_active_2002,mikovaIncreasingEfficiencyLinearoptical2012, meyer-scottSinglephotonSourcesApproaching2020}.

We present a fiber-compatible feed-forward scheme and apply it to remote preparation \cite{bennettRSP} of single-qubit states encoded in the polarization of telecommunication-wavelength photons.
We experimentally implement our scheme using relatively inexpensive, off-the-shelf components compatible with standard telecommunication technologies and thus demonstrate a high-speed and high-fidelity photonic feed-forward protocol.
We verify this by remotely preparing the polarization state of a single photon without post-selection, and we achieve an average fidelity of (99.0 $\pm$ 1)\%, where the  error bar includes statistical errors such as finite photon statistics and waveplate errors, after correcting for imperfections in generating the entangled state.

\begin{figure*}[t]
    \centering\includegraphics[width=2.05\columnwidth]{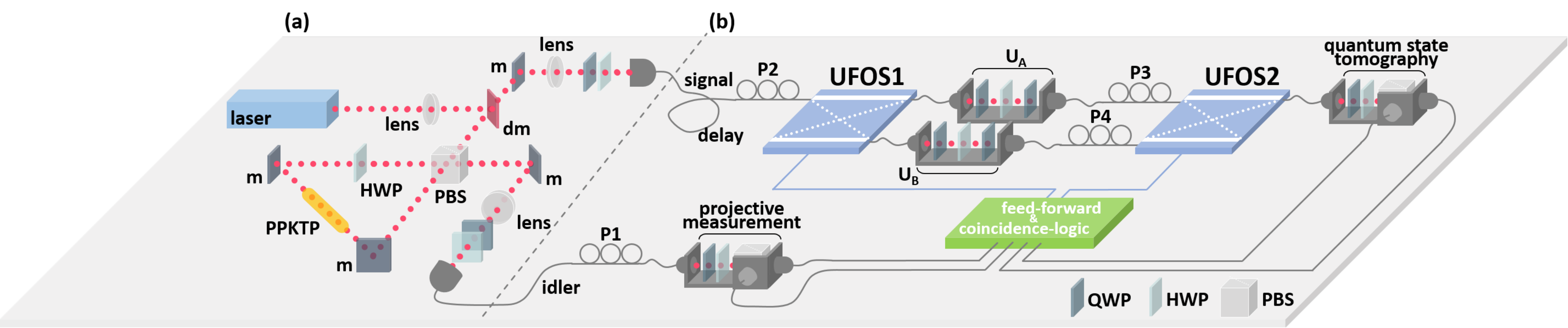}
    \caption{\textbf{(a)} \textit{Type-II SPDC source ---} A $775$ nm, CW beam pumps a PPKTP crystal in a Sagnac-loop configuration, yielding the singlet state $\ket{\psi^-}$.  Here `dm' is a dichroic mirror reflecting $775$ nm light and transmitting $1550$ nm, while `m' are standard mirrors.
    \textbf{(b)} \textit{Feed-Forward setup ---} Photons pairs are sent towards two different stations. A projective measurement is performed on the polarization of the idler photon which is used to deterministically route the sister photon into one of the two paths using classical feed-forward control of ultrafast optical switches (UFOS). Detecting the idler photon also heralds the presence of the sister photon in the other half of the experiment, thus setting a time reference for the UFOS. In the two different paths that the signal photon may take, either operation $U_A$ or $U_B$ (an arbitrary polarization unitary) is performed. Fiber paddles P1-P4 and the waveplates in the source are used to correct for polarization rotations in the optical fiber. The state of the signal photon is finally measured by quantum state tomography.}
    \label{fig:setup}
\end{figure*} 

\section{Fiber-compatible photonic feed-forward}\label{sec:fiberbased}

Our fiber-compatible feed-forward is built around a pair of 2x2 in-fiber ultrafast optical switches, the BATi 2x2 Nanona switch.
These optical switches can route light from two input modes into two output modes with a variable splitting ratio.
The response time of our UFOS is below 60 ns, with a maximal duty-cycle of 1 MHz, and a cross-channel isolation greater than 20 dB for any polarization.
Although a single UFOS controls the path of incident light, we use two UFOSs together with passive polarization optics to implement ultrafast feed-forward operations on the polarization state of single photons, as sketched in Fig.\ref{fig:setup}.
{Note that with two switches we are only able to switch between two operations.  Switching between N different operations could be done, but it would require $\approx \log N$ switches.  For large N, using three Pockels cells to directly implement the feed-forward operation may be less resource intensive.}

To demonstrate our feed-forward protocol we first generate photon pairs using spontaneous parametric down-conversion (SPDC), as sketched in Fig.\ref{fig:setup} \textbf{(a)} and discussed below.
The feed-forward begins by performing a projective measurement on the polarization degree of freedom (DOF) of the idler photon by means of a quarter waveplate (QWP) and a half waveplate (HWP) followed by a polarizing beam splitter (PBS).
Photons at the output of the PBS are coupled to single-mode fibers and detected by two superconducting nanowire single-photon detectors (SNSPD) from PhotonSpot Inc. (deadtime $50$ ns, average system detection efficiency $90\%$).
Detection events are recorded and processed by a commercial time tagging module (UQDevices Logic16 TTM).
The TTM is programmed to generate an output TTL pulse ($5$ V high, and $700$ ns duration) when a photon is detected in the transmitted port of the PBS, and to do nothing otherwise.
This classical electronic feed-forward signal is used to control the UFOS, as indicated by the blue lines in Fig.\ref{fig:setup} \textbf{(b)}.
In the absence of a triggering pulse, the UFOS are set to a ``bar state'': UFOS1 routes the signal photon to unitary transform $U_A$, and UFOS2 sends it towards the quantum state tomography (QST) station.
The triggering pulse sets both UFOS in the ``cross state'' for 700 ns (the duration of the feed-forward pulse).
In this case, UFOS1 routes the signal photon through unitary transform $U_B$, and UFOS2 sends it towards the QST station.
Note that the signal photon exits in the same mode towards the QST stage in both cases.

We implement $U_A$ and $U_B$ by briefly outcoupling to free space in a ThorLabs u-bench system, and then use a HWP sandwiched between two QWPs to implement an arbitrary polarization unitary (see Appendix \ref{app:C}).
The coupling loss of our u-benches is less than $0.7$ dB, which is comparable to the $1.3$ dB insertion loss of the BATi UFOS.
{ Our use of these free-space passive optics is key to our high-fidelity feed-forward. Although they are a free-space optical elements,
the u-benches allow us to use these well-calibrated passive bulk optics to implement $U_A$ and $U_B$ in a fiber-compatible form factor.
Another approach could be to use fiber-paddles, Faraday rotators or other in-fiber polarization rotators \cite{han2016wavelength} to directly implement $U_A$ and $U_B$, but this would likely come at the cost of a lower fidelity.
}
This is in contrast to the standard approach using Pockels cells, where the birefringence is actively modulated to implement the unitary operation{, which is less amenable to fiber-based applications such as communication networks}.
On the other hand, the active components in our work are standard UFOS, which can set the desired path with a crosstalk of less than 20 dB.
{Specifically, the UFOS we use are operated with a half-voltage of 5V (to be compared with the hundred-volt voltage required to operate Pockels cells \cite{gillettExperimentalFeedbackControl2010,heTimeBinEncodedBosonSampling2017}) and a repetition rate of 1MHz. Our method could be made faster by resorting to existing more expensive, faster switches.}
Additionally, we can easily place different passive polarization optics in either path to enact different and arbitrary $U_A$ and $U_B$. Note that we must compensate for the fixed (but random) polarization rotation implemented by our single-mode optical fibers (see Appendix B).
{Fiber-based feed-forward may also be implemented by encoding the qubit in the path DOF instead of the polarization DOF, in which case the UFOS are replaced by other integrated electro-optic modulators that can be as fast as, and operate with a lower half-voltage than, the UFOS --- see \textit{eg} \cite{mikovaIncreasingEfficiencyLinearoptical2012} in which a fidelity of 97.6$\%$ was obtained.}

The delay between detecting the signal photon at the projective measurement (PM) stage and changing the state of the two UFOS is approximately $560$ ns. This consists of: $\approx 160$ ns for the electrical signal from the SNSPD to reach the TTM, $\approx 300$ ns for the TTM to generate the output pulse, and another $\approx 100$ ns for the signal to propagate to the UFOS and change their state.
{ Many of these latencies can easily be improved. For example, the $\approx 300$ ns processing time of our TTM can be pushed down to a few nanoseconds by triggering the switch directly with the detector signal.}
Hence we use a fiber link of $162$ m to transmit and delay the arrival of the signal photon at the UFOS1 (by $\approx 800$ ns) such that enough time has elapsed for the state of the idler photon to be fed-forward to UFOS1.
We further leave both UFOS in the cross state for $700$ ns to give ample time for the signal photon to traverse the feed-forward operations. 

The combination of these various features allows us to implement high-fidelity feed-forward operations with a net loss of only $\approx 3$ dB. We will now discuss how we benchmark our feed-forward by performing deterministic RSP, but our path-based approach to feed-forward could easily be used in a host of photonic quantum information settings.

\section{Remote state preparation}\label{sec:remotestate}
We use the setup shown in Fig.\ref{fig:setup} to implement post-selection--free RSP.
In what follows, $\ket{H}$ ($\ket{V}$) represents a state of linear polarization along the horizontal (vertical) axis, and the idler and signal photons are identified with indices $i$ and $s$, respectively.
At the bottom station, a projective measurement on the polarization degree of freedom of the idler photon is performed: it is projected in either
\begin{align}
        \ket{\Psi}_i=\alpha\ket{H}_i+\beta\ket{V}_i \label{eq:idlerstate}
        \intertext{or}
        \ket{\Psi^\bot}_i=\alpha\ket{V}_i-\beta\ket{H}_i, \label{eq:idlerstateperp}
\end{align}
with $\alpha=\cos{\theta/2}$, and $\beta=\sin{\theta/2}\exp{i\phi}$. These coefficients are controlled by setting the angle of the HWP and QWP before the PBS (see Appendix \ref{app:C}). 
The result of the measurement (\textit{i.e.}, whether the idler photon is in state $\ket{\Psi}_i$ or $\ket{\Psi^\bot}_i$) is sent to the upper station by a classical communication channel.

The signal photon passes through a fiber delay, arriving at the upper station \textit{after} the result of the projective measurement on the idler photon has reached it.
Upon arrival at the upper station, the signal photon has been collapsed into either of the two states
\begin{align}
\label{eq:signalinitialstate}
        \ket{\Psi}_s=\bra{\Psi^\bot}_i\ket{\psi^-}=\alpha\ket{H}_s+\beta\ket{V}_s
        \intertext{or}
        \ket{\Psi^\bot}_s=\bra{\Psi}_i\ket{\psi^-}=\alpha\ket{V}_s-\beta\ket{H}_s,
\end{align}
depending on the outcome of the first projective measurement.

Hence, performing a projective measurement on the polarization DOF of the idler photon sets the polarization of the signal photon into one of two definite states, which we know without having to measure it.
Conditionally on the state of the idler photon, we can either do nothing (which corresponds to applying unitary operation $U_A=\mathbb{1}$) or \textit{correct} the state of the signal photon so that it is always in $\ket{\Psi}$:
\begin{equation}
    \label{eq:signalcorr}
\begin{aligned}
\mathrm{if}
\end{aligned}
\left\{
\qquad
\begin{aligned}
\ket{\Psi}_i&=\ket{\Psi^\bot}_i \rightarrow \mathbb{1}\ket{\Psi}_s=\ket{\Psi}_s, \\
\ket{\Psi}_i&=\ket{\Psi}_i\rightarrow U_B\ket{\Psi^\bot}_s=\ket{\Psi}_s.
\end{aligned}
\right.
\end{equation}
\begin{figure}[t]
    \centering
    \includegraphics[width=1\columnwidth]{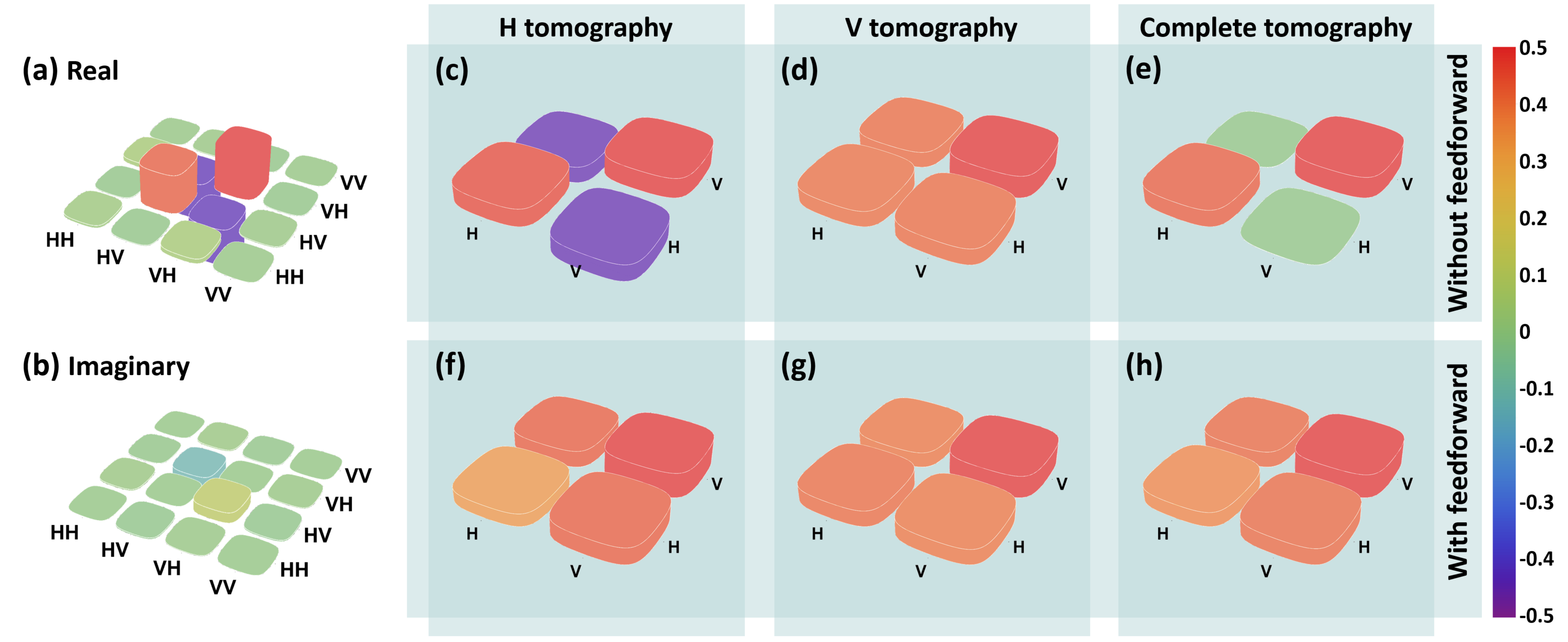}
    \caption{Quantum state tomography of the input and output states. The color bar on the right refers to the amplitude of all the density matrices in panels \textbf{(a)}-\textbf{(h)}.
    Left: \textbf{(a)} Real- and \textbf{(b)} imaginary-parts of the density matrix elements of the two-photon input state for a measurement on the equatorial plane.
    Right: real part of the density matrix elements of the one-photon output states \textbf{(f)-(h)} with and \textbf{(c)-(e)} without feed-forward.
    The imaginary components are relatively small, and not shown.
    The H-tomography (V-tomography) results are post-selected on the idler photon being transmitted (reflected) in the projective measurement (PM). In the complete tomography results no post-selection is performed.}
    \label{fig:instate}
\end{figure}
Note that in general $U_B$ depends on $\theta$ and $\phi$ \cite{pengExperimentalImplementationRemote2003}.
However, there are classes of states for which the $U_B$ is independent of the state one wishes to prepare: this is the standard RSP result \cite{bennettRSP}.
We consider two specific cases in which $U_B=i\sigma_y$ and $U_B=\sigma_z$.

In the first case, $\phi=0$ and the signal's state is prepared on a meridian plane of the Bloch sphere (HDVA plane) --- $\ket{\Psi}_s=\cos{\theta/2}\ket{H}+\sin{\theta/2}\ket{V}$. To produce this family of states we remove the QWP from the PM station and vary the HWP from $0$ to $90\degree$.
In the second case, the signal's state is prepared as an arbitrary equatorial state (DRAL plane) --- $\ket{\Psi}_s=\frac{1}{\sqrt{2}}\left(\ket{H}+e^{i\phi}\ket{V}\right)$. This projection is made by setting the QWP to $45\degree$ and varying the HWP from $22.5$ to $112.5\degree$.

\section{Experiment}\label{sec:experiment}
The input state is generated by type-II spontaneous parametric down conversion, as shown in Fig.\ref{fig:setup} \textbf{(b)}.
The source of entangled photons with a degenerate wavelength is based on a Sagnac loop.
It consists of a PPKTP crystal, with a length of $30$ mm, a poling period of $46.2$ $\mu$m, and tunable temperature.
The crystal is pumped by an $80$ mW, CW laser of wavelength $775$ nm.
Ultra-narrow bandpass filters with a full width at half maximum bandwidth of $3.2$ nm and an almost top-hat shape transmittivity are used to filter the photons.
Directly after the source, we measure singles rates of $85$ kHz and a coincidence rate of $17$ kHz with our SNSPDs, corresponding to a pair coupling efficiency of $20\%$.
We use the same TTM to measure the count rates and generate the feed-forward signal.
The idler and signal photons are sent to their respective stations by fiber (SMF28) links.
After coupling through our experiment, we measure singles rates of $50$ kHz and $30$ kHz at the PM and QST stations, respectively, and a net coincidence rate of $\approx 3$ kHz.

\begin{figure}[t]
    \centering
    \includegraphics[width=0.97\columnwidth]{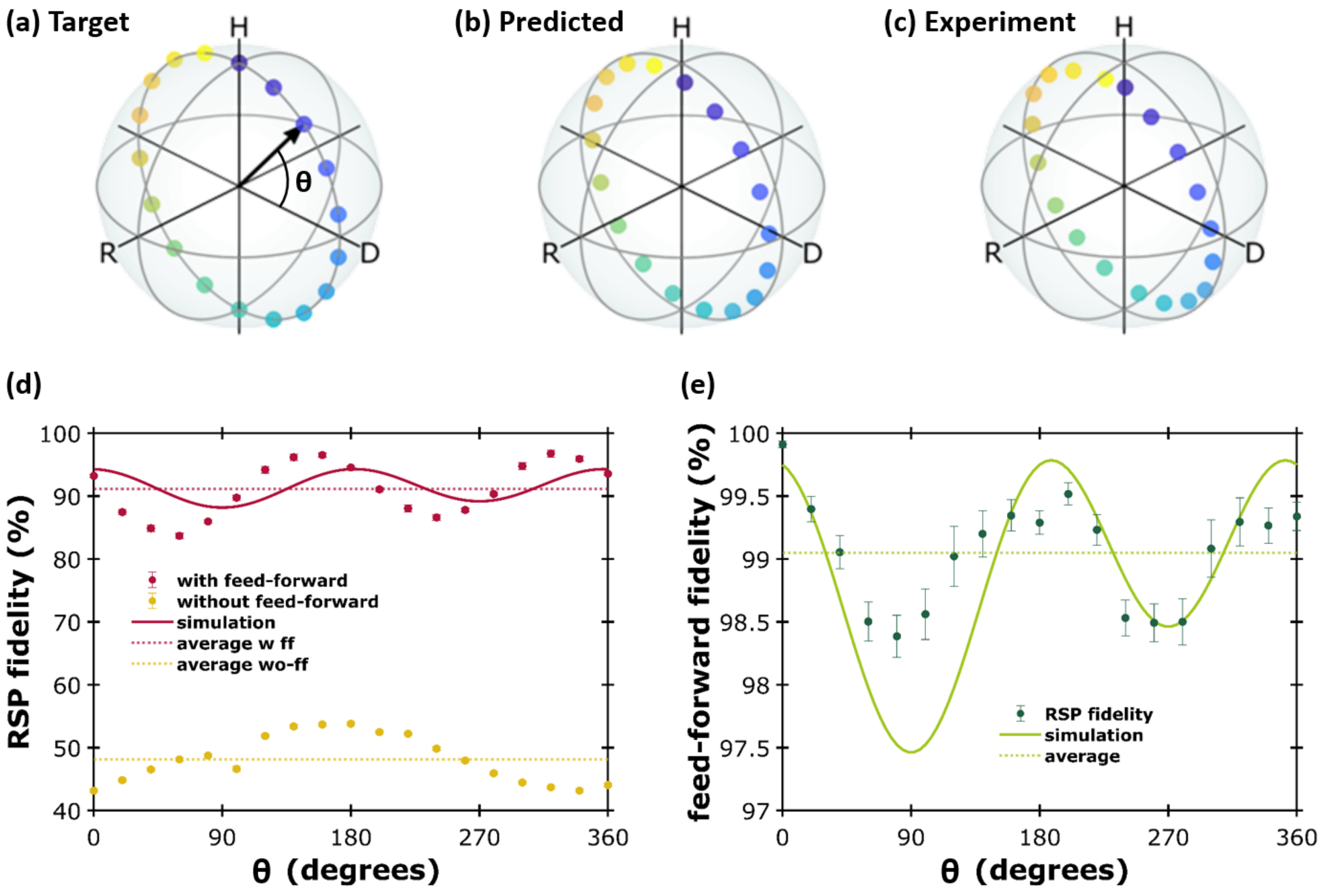}
    \caption{Preparation of the signal photon on the meridian plane of the Bloch sphere. \textbf{(a)} Target signal states, assuming a perfect singlet state. \textbf{(b)} Predicted signal states theoretically calculated accounting for experimental imperfections in generating the singlet state. \textbf{(c)} Measured signal states. The different colors of the points in panels \textbf{(a)}-\textbf{(c)} correspond to the same experimental setting.
    \textbf{(d)} Fidelity of the measured experimental states with the targets with (red) and without (yellow) feed-forward plotter versus the Bloch angle of the target state (black arrow in \textbf{(a)}). This Bloch angle maps onto the half waveplate angle in the protective measurement station as $\theta=4\theta
   `$. 
   \textbf{(e)} Fidelity of the experimentally prepared states with the predicted states versus the Bloch angle.  \label{fig:projreal}}
\end{figure}

In Fig.\ref{fig:instate} \textbf{(a)} and \textbf{(b)}, we show the real and imaginary parts of the reconstructed density matrix of the two-photon entangled state obtained by using quantum state tomography based on a least-squares optimization \cite{qi2013quantum} at the output of the setup (using the PM and QST stations) when the feed-forward is turned off; see Appendix A for details.
{For these measurements, we acquire $\approx 40$,$000$ two-photon events per measurement basis.
We measure typical fidelities of ($92\pm 1$)$\%$  to the Bell singlet state $\ket{\psi^-}$ and purities of ($89\pm2$)$\%$, where the error bars are estimated by performing a Monte Carlo simulation starting with the experimental data, and then adding the Poissonian fluctuations from $40$,$000$ counts, and uncertainties of $0.5$ degrees in setting the waveplate angles.}

The decreased fidelity from the perfect Bell state is mostly due to distinguishability between the two pathways in the Sagnac source, and a limited amount of polarization rotation in the optical fibers.
This decreased fidelity will also result in a lower RSP fidelity. However, we stress that this decrease is distinct from any errors introduced by our feed-forward. Hence, we will use these two-qubit tomography results to treat these error sources separately (see Appendix \ref{app:A}).

We first measure the net RSP fidelity, by projecting the idler photon into a given basis and then estimating a series of single-qubit density matrices at the QST station.

The plots in the shaded area on the right of Fig.\ref{fig:instate} (showing the real part of the reconstructed density matrices with (\textbf{(f)-(h)}) and without (\textbf{(c)-(e)}) feed-forward) illustrate the effect of our protocol on the signal photon's state.
In these data, the idler photon is measured in the $(\ket{H}\pm\ket{V})/\sqrt{2}$ basis, and each row shows three different analyses of the results in the same configuration.
The `H-tomography' (panels \textbf{(c)} and \textbf{(f)}), corresponds to events where we post-select on the idler photon being transmitted through the PM beamsplitter (\textit{i.e.} it is found to be $(\ket{H}+\ket{V})/\sqrt{2}$).  Without feed-forward, this collapses the signal photon into $(\ket{H}-\ket{V})/\sqrt{2}$ (since we began with a singlet state); this is evident in negative off-diagonal components of the density matrix of panel \textbf{(c)}.
However, when the feed-forward is enabled (panel \textbf{(f)}) the state is flipped to $(\ket{H}+\ket{V})/\sqrt{2}$, and the off-diagonal components become positive.
Similarly, the `V-tomography' results come from events where the idler photon is reflected, projecting the signal photon into $(\ket{H}-\ket{V})/\sqrt{2}$. In this case, the data with \textbf{(g)} and without \textbf{(d)} feed-forward agree, since the feed-forward operation is applied.
Finally, the complete tomography results used no post-selection. Hence, without feed-forward the signal photon is in a mixed state (the off-diagonal components in panel \textbf{(e)} are $\approx 0$), but using feed-forward allows us to deterministically prepare $(\ket{H}+\ket{V})/\sqrt{2}$.

\begin{figure}[t]
    \centering
    \includegraphics[width=0.97\columnwidth]{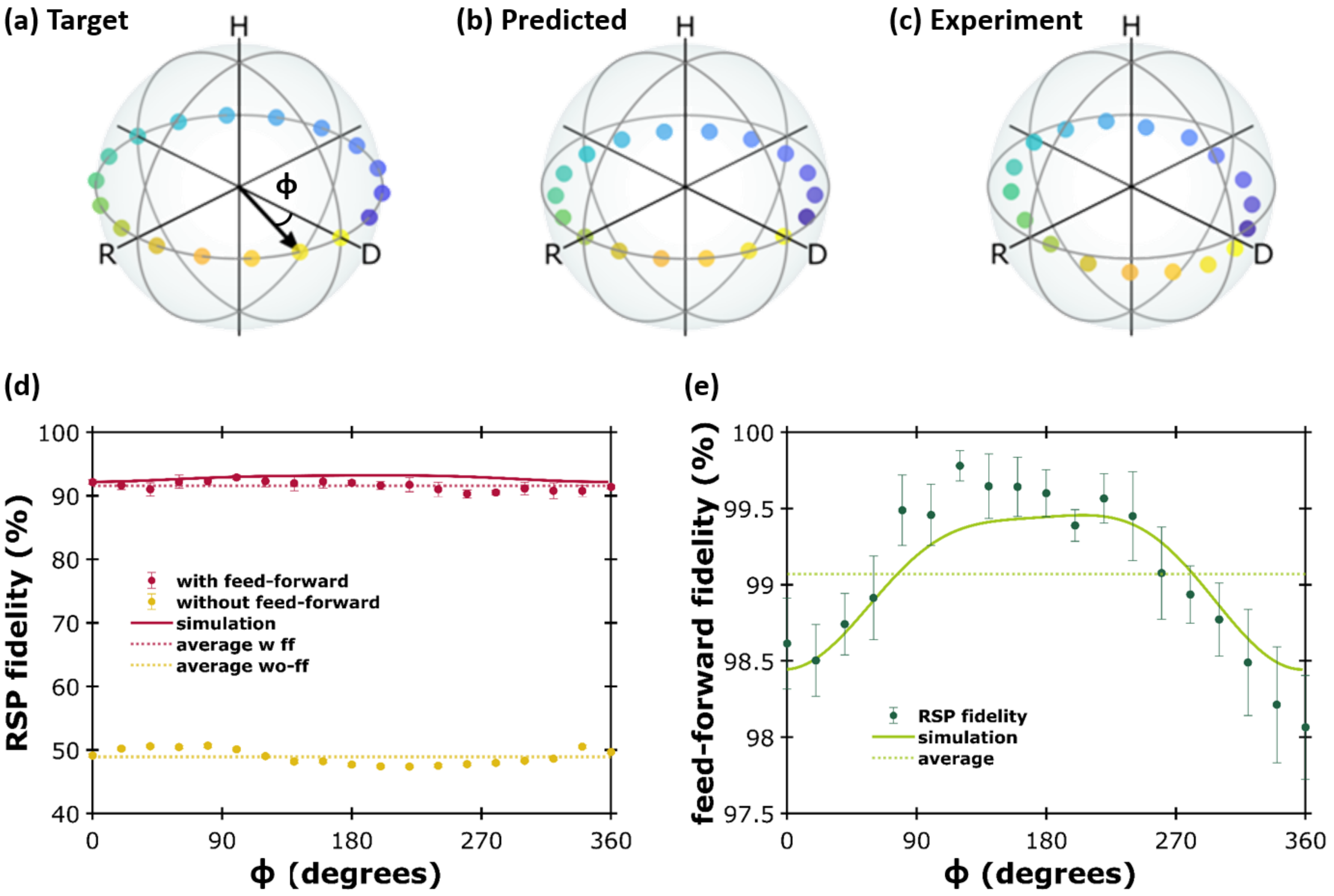}
    \caption{Preparation of the signal photon on the equatorial plane of the Bloch sphere. The data in all panels are plotted as in Fig. \ref{fig:projreal}, except, here, $\phi$ of the target states in the equatorial plane is related to the half waveplate angle $\theta'$ as $\phi=4(\theta'-22.5)$.
    \label{fig:projcplx}}
\end{figure}

To further characterize the RSP fidelity, we prepare a range of states along the meridian and equatorial planes, and perform QST on the output{, obtaining approximately $35$,$000$ photon counts per measurement setting.}
The target states (\textit{i.e.} the states we attempt to remotely prepare) are plotted as points on the Bloch sphere in panels \textbf{(a)} of 
Fig.\ref{fig:projreal} and Fig.\ref{fig:projcplx}, and the resulting experimental states are plotted in panels \textbf{(c)} of the same figures.
The fidelities between the experimental states and the target states with the feed-forward turned off are plotted in yellow in panels (\textbf{d}), and have an average fidelity of (49 $\pm$ 2)\% to the target state. The uncertainty is obtained by propagating the individual uncertainties.
{The error bars presented in panels \textbf{(c)} and \textbf{(d)} of Figs.\ref{fig:projreal} and \ref{fig:projcplx} are calculated from a Monte Carlo simulation including a 0.5 degree error on setting the tomography waveplates, and the finite counting statistics.}
This low fidelity is to be expected, since, in the absence of feed-forward, the signal photon is a maximally mixed state and we have not post-selected our results.
When the feed-forward is enabled (in red in panels \textbf{(d)}), the fidelity increases dramatically.
For these data, we obtain average RSP fidelities of (91 $\pm$ 1)\% and (92 $\pm$ 1)\% to the target state when projecting on the equatorial and meridian planes, respectively. 
However, these fidelities are severely limited by the reduced fidelity of the experimental entangled state with $\ket{\psi^-}$.
Again, we stress that the reduction of the fidelity of our entangled state comes primarily from two sources. The first is from distinguishability between the HV and VH paths in the Sagnac interferometer, which reduces fidelity of the entangled state (by entangling to other degrees of freedom, hence reducing the purity).
The second error source is imperfect polarization compensation in our fiber network (see Appendix B).

{We model both of these imperfections, and the result is plotted as the red line in panel (d) of the two figures.
Our model also includes a $1\%$ polarization dependent loss, which is consistent with the specified $<0.2$ dB polarization dependent loss of our switches.
Although it is difficult to exactly determine the form of the imperfect polarization compensation, we find that that a residual birefringence of $0.5$ rad on one side of the experiment is sufficient to describe the oscillations observed in the data presented in Fig.\ref{fig:projreal}, while a birefringence of $0.25$ rad fits well the data in Fig.\ref{fig:projcplx} \textbf{(d)}.  (Since the polarization compensation was redone between these two measurements, it is reasonable to have different values for this.)
The simulations presented in both figures assume a purity of the $0.89$ for the shared Bell state.
}

To assess the feed-forward fidelity directly, we use the two-photon tomography results presented above to infer and remove the errors due to the state generation.
To do this, we predict the states we expect to remotely prepare given our actual experimental entangled state.  
These predicted states are plotted in panels \textbf{(b)} of Fig.\ref{fig:projreal} and Fig.\ref{fig:projcplx}. The lower-than-unity fidelity between these states and the corresponding experimental states are caused by imperfections in our feed-forward. 
{ These errors come from a combination of slightly miscalibrated waveplates used to implement $U_A$ and $U_B$, polarization dependent losses in the switch (which our correction does not account for), and imperfect polarization compensation.
The simulations presented in panels (\textbf{e}) of both figures assume a perfect singlet state, polarization dependent losses of $1\%$, and residual birefringence of $0.2$ rad.
}
Leakage into the incorrect path could cause additional errors, but we estimate this to be negligible.
The resulting feed-forward fidelities are plotted in green in panels \textbf{(e)} of both figures. The average feed-forward fidelity of these data is (99.0 $\pm$ 1)\% and (99.0 $\pm$ 1)\%, for states in the meridian and equatorial planes, respectively.
This demonstrates that our feed-forward indeed operates with extremely high-fidelity,
{ one of the highest in the state-of-the-art to the best of our knowledge. Our low error rate is largely due to our use of well-calibrated passive optics, which could in fact be implemented with a variety of other switching technologies, also in bulk or integrated settings.
Furthermore, the switches and electronics used in our experiment were not state-of-the-art, and the timing performance of our techniques could easily be improved with either faster commercial or research-level switches.
}

\section{Discussion}
Our feed-forward scheme enables the remote state preparation of the polarization state of single photons without post-selection.
In our photonics implementation we use a novel technique for fiber-compatible feed-forward, exploiting the precision of passive polarization optics in combination with high-quality ultrafast optical switches.
We find that the errors introduced into remote state preparation by our feed-forward operation are less than 1$\%$.
This figure of merit, combined with the use of standard, off-the-shelf photonic technology for telecommunication-wavelengths, makes our methods usable for quantum photonics tasks in which fast switching and routing of light are critical. 
{The experimental platform could be improved by resorting to faster switches realized with \textit{e.g.} four-wave mixing in interferometers, which promises rates up to 500 MHz with losses below 1 dB \cite{leeLowLossHighSpeedFiberOptic2018}.}
{Applications of our methods include}: measurement based quantum computation \cite{saggio_experimental_2019}, experiments on the foundations of quantum physics \cite{maclean_quantum-coherent_2017,saunders_experimental_2017,vedovato_postselection-loophole-free_2018}{, neuromorphic computing}\cite{shen_deep_2017} or quantum information applications like photonic simulations \cite{schreiber_2d_2012}, photon counting \cite{tiedau_high_2019}, or establishing a quantum internet \cite{kimble2008quantum}.


\section*{Acknowledgements}
The authors are thankful to Valeria Saggio, Bob Peterson and Teo Str\"omberg for insightful discussions.

\section*{Funding}
Conselho Nacional de Desenvolvimento Científico e Tecnológico (204937/2018-3); Red Bull GmbH; Air Force Office of Scientific Research (QAT4SECOMP (FA2386-17-1-4011)); Austrian Science Fund (FWF): F 7113-N38 (BeyondC), FG 5-N (Research Group), P 30817-N36 (GRIPS) and P 30067\_N36 (NaMuG);
{\"O}sterreichische Forschungsf{\"o}rderungsgesellschaft (QuantERA ERA-NET Cofund project HiPhoP (No.731473)); 
Research Platform for Testing the Quantum and Gravity Interface (TURIS), the European Commission (ErBeSta (No.800942)), Christian Doppler Forschungsgesellschaft; {\"O}sterreichische Nationalstiftung f{\"u}rForschung, Technologie und Entwicklung; Bundesministerium f{\"u}r Digitalisierung und Wirtschaftsstandort.

\section*{Disclosures} 
The authors declare no conflicts of interest.
\newline


\appendix
\section*{Appendix}

\section{Quantum state tomography}\label{app:A}

In order to estimate the feed-forward fidelity we use both the results of two-photon quantum state tomography (QST) on the shared entangled state, and single-photon QST on the remotely prepared states. 
For both of these analyses we use a least-squares fitting routine to reconstruct the two- and one-photon density matrices.
The single-qubit measurements are carried out in the QST station (see Fig. \ref{fig:setup}) using a QWP, HWP and PBS mounted in a fiber u-bench.
For the `complete tomography results' used to assess the feed-forward fidelity, we measure counts in the transmitted and reflected ports of the QST station in coincidence with either of the two detectors in the projective measurement (PM) station.
The waveplate angles used are presented in the first column of Table \ref{tab:tableangles}.
For the two-photon tomography, we set both UFOS in the bar state, so the signal photon is transmitted through $U_A$ (which implements identity).
We then measure all four coincidence rates between the two detectors in the QST station and the two detectors in the PM station.
Setting the waveplates to the 9 settings presented in the second column of Table \ref{tab:tableangles} yields an over-complete measurement set of 36 coincidence rates.
From these measurements, the probability of projecting the two-photon state $\rho$ onto the corresponding projectors is estimated.

\begin{table}[h]\setlength{\arrayrulewidth}{0.3mm}
\begin{center}
\begin{scriptsize}
	\begin{tabular}[t]{ |c|c|c| }
      \hline
      \hline  
            \thead{Single-Photon \\ Measurement Set} & \multicolumn{2}{c|}{\thead{Two-Photon \\State Tomography}}    \\
             \hline
             (1) $0\degree_{HWP}$, $0\degree_{QWP}$ 
             & QST   
             & PM 
             \tabularnewline
             \cline{2-3}
             (2) $0\degree_{HWP}$, $45\degree_{QWP}$ 
             & (1) $0\degree_{HWP}$, $0\degree_{QWP}$ 
             & $0\degree_{HWP}$, $0\degree_{QWP}$
			\tabularnewline
             (3) $22.5\degree_{HWP}$, $45\degree_{QWP}$
             &(2) $0\degree_{HWP}$, $0\degree_{QWP}$ 
             & $22.5\degree_{HWP}$, $45\degree_{QWP}$
			\tabularnewline
             &(3) $0\degree_{HWP}$, $0\degree_{QWP}$ 
             & $0\degree_{HWP}$, $45\degree_{QWP}$
             \tabularnewline 
             &  (4) $22.5\degree_{HWP}$, $45\degree_{QWP}$ 
             &$0\degree_{HWP}$, $45\degree_{QWP}$
             \tabularnewline 
              &  (5) $22.5\degree_{HWP}$, $45\degree_{QWP}$ 
              &$22.5\degree_{HWP}$, $45\degree_{QWP}$
              \tabularnewline 
              &   (6) $22.5\degree_{HWP}$, $45\degree_{QWP}$ 
              & $0\degree_{HWP}$, $45\degree_{QWP}$
             \tabularnewline 
              &   (7) $0\degree_{HWP}$, $45\degree_{QWP}$ 
              & $0\degree_{HWP}$, $0\degree_{QWP}$
             \tabularnewline 
              &   (8) $0\degree_{HWP}$, $45\degree_{QWP}$ 
              & $22.5\degree_{HWP}$, $45\degree_{QWP}$
            \tabularnewline 
              &   (9) $0\degree_{HWP}$, $45\degree_{QWP}$ 
              & $0\degree_{HWP}$, $45\degree_{QWP}$
 			\tabularnewline 
 			\hline
	\end{tabular}
\end{scriptsize}
\caption{Set of waveplate angles used to perform the single-photon and two-photon quantum state tomography. \label{tab:tableangles}}
\end{center}
\end{table} 

For example, for the two-photon measurement with all of the waveplates set to $0\degree$, the total number of events wherein both photons are transmitted ($C_{HH}$), both photons are reflected ($C_{VV}$), one transmitting and one reflecting ($C_{HV}$), and vice versa ($C_{VH}$) are measured.
From these counts we estimate probabilities $P_{HH}$, $P_{HV}$, $P_{VH}$, and $P_{VV}$ as $P_{HH}=C_{HH}/(C_{HH}+C_{HV}+C_{VH}+C_{VV})$, and so on.
These probabilities represent the probability to project the experimentally prepared state onto the projection operator $\ket{H}\bra{H}\otimes\ket{H}\bra{H}$, \textit{etc}.
We repeat this procedure for the 9 measurement settings listed in Table. \ref{tab:tableangles}.

Finally, we construct a least-squares estimate of the density matrix $\tilde{\rho}_\mathrm{LS}$ by minimising the following cost function
\begin{equation}
\label{eq:ls}
C=\sum_{s,i} \left( \mathrm{Tr}\left[ \hat{O}_{s,i} \hat{\rho}_\mathrm{LS} \right] - P_{s,i} \right)^2
\end{equation}
over $\hat{\rho}_\mathrm{LS}$ using SeDuMi and YALMIP in MATLAB. In the above equation,
\begin{equation}
P_{s,i}= \left\{P_{HH}, P_{HV}, P_{VH}, P_{VV}, P_{DD}, \dots \right\},
\end{equation}
and
\begin{equation}
\hat{O}_{s,i}=\left\{\ket{HH}\bra{HH},\ket{HV}\bra{HV}, \dots \right\},
\end{equation}
where $\ket{HH}\bra{HH}$ is short for $\ket{H}\bra{H}\otimes\ket{H}\bra{H}$.
Furthermore, $\hat{\rho}_\mathrm{LS}$ is constrained to be a positive Hermitian matrix of trace 1, which is ensured by the MATLAB packages used.

For the single-photon tomography we perform essentially the same routine, but with single-photon projectors and probabilities.

\section{Fiber polarization compensation}\label{app:B}

The experimental setup (Fig.\ref{fig:setup} \textbf{(b)}) is composed of single-mode optical fibers (SMF28) from which the light is outcoupled only briefly in stable u-bench systems to implement the single-qubit measurements and unitary transforms on the polarization of the photon.
Standard optical fibers apply a fixed, but random polarization transformation, which must be corrected (or compensated) in order to transmit quantum information in the polarization degree-of-freedom.
For our feed-forward to function correctly, this means that we must ensure that each leg of optical fiber implements the identity operation.
We do this with a combination of waveplates and fiber paddles as follows.

In order to correct the fiber connecting the source to the projective measurement (PM) station we insert a calibrated free-space polarizer in the idler path of the source before the final QWP in front of the fiber coupler.
This allows us to prepare linear polarization in $\ket{H}$ and $\ket{D}$.
By ensuring that both $\ket{H}$ and $\ket{D}$ photons emerge from the fiber in the same state they entered the fiber, we set the fiber to effectively implement identity on the polarization.

To this end, we set the QWP to $45\degree$ to decouple the $\{\ket{D},\ket{A}\}$ basis from the $\{\ket{H}/\ket{V}\}$ basis. 
We then use the paddles P1 to compensate the polarization in the $\{\ket{H},\ket{V}\}$ basis, and the final HWP in the source to compensate in the $\{\ket{D},\ket{A}\}$ basis.
To do so, in the $\{\ket{H},\ket{V}\}$ basis ($\{\ket{D},\ket{A}\}$ basis) we set the PM to measure in $\{\ket{H},\ket{V}\}$ ($\{\ket{D},\ket{A}\}$) and use the paddles (HWP) to minimize the counts in the transmitted port of the PM.
This procedure is repeated iteratively, until the counts in the transmitted port of the PM are satisfactorily minimized in both bases.
We then remove the polarizer placed inside the source.

The signal photon's path is more complex, as there are three individual legs in the RSP station that must be treated separately.
Similarly to the idler photon, we first insert a polarizer in the source before the final QWP, and use that QWP at $45\degree$ to decouple the  $\{\ket{H},\ket{V}\}$ basis from the $\{\ket{D},\ket{A}\}$ basis.
We begin with the leg from the source to the $U_B$ u-bench by setting the UFOS to the cross state, directing light down, through $U_B$.
Afterwards, we remove the waveplates from the $U_B$ u-bench and insert a polarizer, which we rotate to measure light in $\{\ket{H},\ket{V}\}$ and $\{\ket{D},\ket{A}\}$ bases.
Then, as we did for the idler photon, we use the HWP and paddles P2 to compensate the polarization, by minimising the transmission through the polarizer in the u-bench (which we set perpendicular to the polarizer in the source) in each basis.

After this, we move on to the fiber connection from the $U_B$ u-bench to the quantum state tomography (QST) station.
We first remove the polarizer from the source, use the QST station to measure in the $\{\ket{H},\ket{V}\}$ and $\{\ket{D},\ket{A}\}$ bases, and use the polarizer in the u-bench to prepare $\ket{H}$ and $\ket{D}$. In this case we do not use a QWP to decouple the two bases so we must iterate, using only the paddles P4 after the $U_B$ u-bench to compensate the polarization in both bases.

The remaining leg is from the source, through the $U_A$ u-bench, to the QST station.
Since we always set $U_A$ to identity, we compensate for this entire path at once.
We thus set the UFOS to the bar state and then use the same technique as the second leg in the $U_B$ path: inserting a polarizer in the source and iterating with the paddle P3 until we are satisfied with both bases.
Once the polarization is compensated, it remains stable for several days.

\textbf{\begin{table}[ht]\setlength{\arrayrulewidth}{0.3mm}
\begin{center}
\begin{scriptsize}
	\begin{tabular}[t]{ |c|c|c| }
      \hline
      \hline
		 \thead{Projected Family\\ of States} & \thead{$U_B$} & \thead{Wave-plate\\Angles for $U_B$}
             \tabularnewline 
             \hline
             $\ket{\Psi}=\cos{\theta /2}\ket{H}+\sin{\theta /2}\ket{V}$  & $i\sigma_{y}$ & $90\degree_{QWP}$, $45\degree_{HWP}$, $0\degree_{QWP}$
			\tabularnewline
             $\ket{\Psi}=\ket{H}+\exp{i\phi}\ket{V}$  & $\sigma_{z}$ & $90\degree_{QWP}$, $0\degree_{HWP}$, $0\degree_{QWP}$
 			\tabularnewline 
 			\hline
	\end{tabular}
\end{scriptsize}
\caption{Set of waveplate angles used to implement the unitary transformation $U_B$ on the signal photon.}\label{tab2}
\end{center}
\end{table}
}

\section{Additional waveplate settings}\label{app:C}

We use the waveplates in the PM station to project the idler photon and remotely prepare the signal photon in the state $\alpha\ket{H}+\beta\ket{V}$. As described in the main text, we choose two families of states on the meridian and equatorial plane of the Bloch sphere.
To project on the meridian plane, we remove the QWP from the PM station. In this case, the angle $\theta$ in the meridian plane of the Bloch sphere (see Fig. \ref{fig:projreal} \textbf{(a)}) can be related to the angle $\theta'$ of the HWP by the following expression:
\begin{equation}
\label{eq:relationthetathetaprime}
 \theta={4}{\theta'}.
\end{equation}
This allows us to prepare states with $\alpha=\cos{\theta/2}$ and $\beta=\sin{\theta/2}$.
For the measurements along the equator, we insert the QWP and fix it to $45\degree$. 
The angle $\phi$ in the equatorial plane of the Bloch sphere (see Fig. \ref{fig:projcplx} \textbf{(a)}) is related to $\theta`$ of the HWP by
\begin{equation}
    \label{eq:relationphithetaprime}
    \phi=4(\theta'-22.5\degree).
\end{equation}
In this configuration we can produce states with $\alpha=\frac{1}{\sqrt{2}}$ and $\beta=\frac{e^{i\phi}}{\sqrt{2}}$.

Each class of states requires a different correction operation to be applied using measurement and feed-forward when the idler photon is transmitted through the PBS in the PM station. For the meridian (equatorial) states we must apply $i\sigma_y$ ($\sigma_z$).
We implement these operations using a HWP sandwiched between two QWP in the $U_B$ u-bench. The specific angles we use are presented in Table \ref{tab2}.

\end{document}